\documentclass{emulateapj}
\usepackage{psfig}
\usepackage{lscape}

\def\arcmin{\hbox{$^\prime$}}

\def\msun{M$_\odot$}

\slugcomment{}

\shorttitle{SN 2007if}
\shortauthors{Yuan et al.}

\begin{document}


\title{The Exceptionally Luminous Type Ia Supernova 2007if}



\author{F.~Yuan\altaffilmark{1}
,~R.~M.~Quimby\altaffilmark{2}
,~J.~C.~Wheeler\altaffilmark{3}
,~J.~Vink\'o\altaffilmark{3}
,~E.~Chatzopoulos\altaffilmark{3}
,~C.~W.~Akerlof\altaffilmark{1}
,~S.~Kulkarni\altaffilmark{2}
,~J.~M.~Miller\altaffilmark{1}
,~T.~A.~McKay\altaffilmark{1}
,~F.~Aharonian\altaffilmark{4,5}}


\altaffiltext{1}{Physics Department, University of Michigan, Ann Arbor, MI 48109}
\altaffiltext{2}{Astronomy Department, California Institute of Technology, 
        105-24, Pasadena, CA 91125}
\altaffiltext{3}{Department of Astronomy, University of Texas, Austin, TX
        78712}
\altaffiltext{4}{Max-Planck-Institut f\"{u}r Kernphysik, Saupfercheckweg 1,
        69117 Heidelberg, Germany}
\altaffiltext{5}{Dublin Institute For Advanced Studies, 31 Fitzwilliam Place, Dublin 2, Ireland}


\begin{abstract}
SN 2007if was the third over-luminous SN Ia detected after 2003fg and 2006gz. We present the photometric and spectroscopic observations of the supernova and its host by ROTSE-III, HET and Keck. From the H$_\alpha$ line identified in the host spectra, we determine a redshift of 0.0736. At this distance, the supernova reached an absolute magnitude of -20.4, brighter than any other SNe Ia ever observed. If the source of luminosity is radioactive decay, a large amount of radioactive nickel ($\sim$1.5~\msun) is required to power the peak luminosity, more than can be produced realistically in a Chandransekar mass progenitor. Low expansion velocity, similar to that of 2003fg, is also measured around the maximum light. The observations may suggest that SN 2007if was from a massive white dwarf progenitor, plausibly exploding with mass well-beyond 1.4~\msun. Alternatively, we investigate circumstellar interaction that may contribute to the excess luminosity.

\end{abstract}



\section{Introduction}
Type Ia supernovae (SNe Ia) are commonly understood to be thermonuclear explosions of carbon/oxygen white dwarfs (CO WDs). In the most likely scenario, a single WD approaches its Chandrasekhar mass limit by accreting from a non-degenerate binary companion (single-degenerate (SD) model, \citealp{1973ApJ...186.1007W, 1982ApJ...253..798N}). The other probable channel is two WDs that merge due to gravitational radiation (double-degenerate (DD) model, \citealp{1984ApJS...54..335I, 1984ApJ...277..355W}). In either case, the explosion results in the total obliteration of the entire star (although see \citealp{1985A&A...150L..21S,1991ApJ...367L..19N} for an alternative outcome of accretion-induced collapse to a neutron star). A range of peak luminosities have been observed for SNe Ia, which is believed to be directly related to the amount of radioactive $^{56}$Ni synthesized \citep{1982ApJ...253..785A}. A number of factors can potentially affect the scatter, e.g. the progenitor age \citep[e.g. ][]{1996AJ....112.2391H,2001ApJ...554L.193H,2006ApJ...648..868S,2008ApJ...685..752G} and metallicity \citep[e.g. ][]{2000AJ....120.1479H,2003ApJ...590L..83T,2009ApJ...691..661H}. Geometric effects, such as an asymmetric ignition \citep{ssrh07, 2008ARA&A..46..433W, 2010ApJ...710..444H}, may at least account for part of the dispersion.

Although a large fraction of SN Ia seem to occupy a relatively small parameter space that allows parameterizing of their peak luminosities based on one or two light curve and spectral characteristics \citep{npbbh95,plssh99,bcmta05,bdb09}, the number of exceptional events is rising with the increasing number of well-observed SNe. Among these, two recent cases, SN 2003fg (also known as SNLS-03D3bb, \citealp{hsnec06}) and SN 2006gz \citep{hgpbd07} brought attention to the possibility of an explosion that exceeded the Chandrasekhar mass. Although an asymmetric explosion model offers an alternative explanation \citep{hsr07}, SN 2003fg was so luminous that it is difficult to explain by geometric effects alone. Such extreme cases might be intrinsically rare, but especially valuable for constraining the nature of the progenitors, and helping to understand how to reliably use SNe Ia as tools to measure the cosmic expansion history for a larger distance range. 

We present in this paper a more recent addition to this category, SN 2007if. The supernova was discovered independently by the 0.45m ROTSE-IIIb telescope at McDonald Observatory, Texas, on Aug 19.28 UT and by the Nearby Supernova Factory (SNfactory) on Aug 25.4 UT \citep{atel1212,cbet1059}. At redshift z$\sim$0.074, SN 2007if reached a peak absolute magnitude exceeding -20 and is one of the most luminous Type Ia supernova ever observed. The exceptional brightness and the spectral features might be consistent with a super-Chandrasekhar mass explosion, but we consider other possibilities. 

In the following sections, we first summarize the photometric and spectroscopic observations of SN 2007if and its host in $\S$ 2. We then discuss the possible interpretation of the observed features in $\S$ 3. This is followed by our conclusions in $\S$ 4. Throughout the paper, we assume a standard cosmology model with the Hubble parameter H$_{\rm 0}$=70~km~s$^{-1}$ Mpc$^{-1}$ and the density parameters $\Omega_{\rm m}$=0.3, $\Omega_{\Lambda}$=0.7. All quoted errors are 1-sigma (68\% confidence), unless otherwise stated.

\section{Observations and Analysis}
\subsection{Photometric Observations by ROTSE-III}
ROTSE-IIIb (at the McDonald Observatory) and ROTSE-IIIc (at the H.E.S.S. site at Mt. Gamsberg, Namibia) monitored the field of SN 2007if on a daily basis, weather permitted. The supernova was observed above detection threshold between August 16.29 UT and December 5.08 UT in 2007. The ROTSE-III images were bias-subtracted and flat-fielded by the automated pipeline. Initial object detections were performed by SExtractor \citep{ba96}. The images were then processed with our custom RPHOT photometry program based on the DAOPHOT \citep{stetson87} psf-fitting photometry package \citep{qryaa06}. The host galaxy, with a R-band magnitude of 22.7 (see $\S$2.3), is well below ROTSE-III's detection limit and does not affect our photometry of the supernova.

The unfiltered thinned ROTSE-III CCDs have a peak sensitivity in the R band wavelength range. We estimate the magnitude zero-point by obtaining the median offset from well-measured SDSS r-band magnitudes of selected field sources. The color distribution of these references covers almost the entire SN Ia color space. The standard deviation of the offsets is about 0.20 magnitude. This value provides an estimate of our zero-point uncertainty. 

To find the light curve maximum, the ROTSE-III data were fit with a light curve template, that is constructed from spectral templates \citep{2007ApJ...663.1187H} and by weighing the g, r and i band light curves with the approximate ROTSE-III CCD efficiency curve. The rise and decay phases are fit with two different stretch factors for two reasons. First, the best fit stretch factor is smaller during the rise (1.12$\pm$0.05) than during the decay (1.53$\pm$0.03). \citet{hayden10} have also noticed that the rise and decay stretches are not necessarily correlated. Second, a typical SN Ia becomes progressively redder just after the maximum. Although the color of SN 2007if is not well constrained by our data, there is no evidence that it is not following a similar trend. Fitting the rise and decay separately minimizes the effect of color evolution on the determination of the light curve peak. Due to the uncertainties in constructing the unfiltered template, the errors in our light curve fitting are likely under-estimated.

\begin{figure}[h]
\centerline{\psfig{file=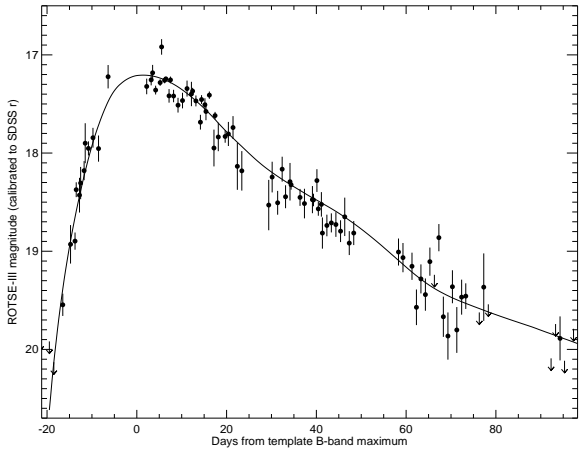,width=3.3in,angle=0}}
\caption{Unfiltered optical light curve of SN 2007if, observed by ROTSE-IIIb and ROTSE-IIIc. The fitted template (\citealp{2007ApJ...663.1187H}, see text) is overplotted as a solid line. \label{fig:2007if_rotse_lc}}
\end{figure}

The best fit maximum date (relative to a B-band template) is found to be September 1.8 UT, with an error of about 0.9 day (90\% confidence, but not including uncertainty from possible color effects). The explosion date is estimated to be around August 9.4 UT, assuming a pre-stretch template rise time of 19.5 days. Our fitting is constrained by an upper limit at August 14.3 UT and the 22-days rise in the rest frame is among the typical range for SNe Ia \citep{hayden10}.

\begin{figure}[h]
\centerline{\psfig{file=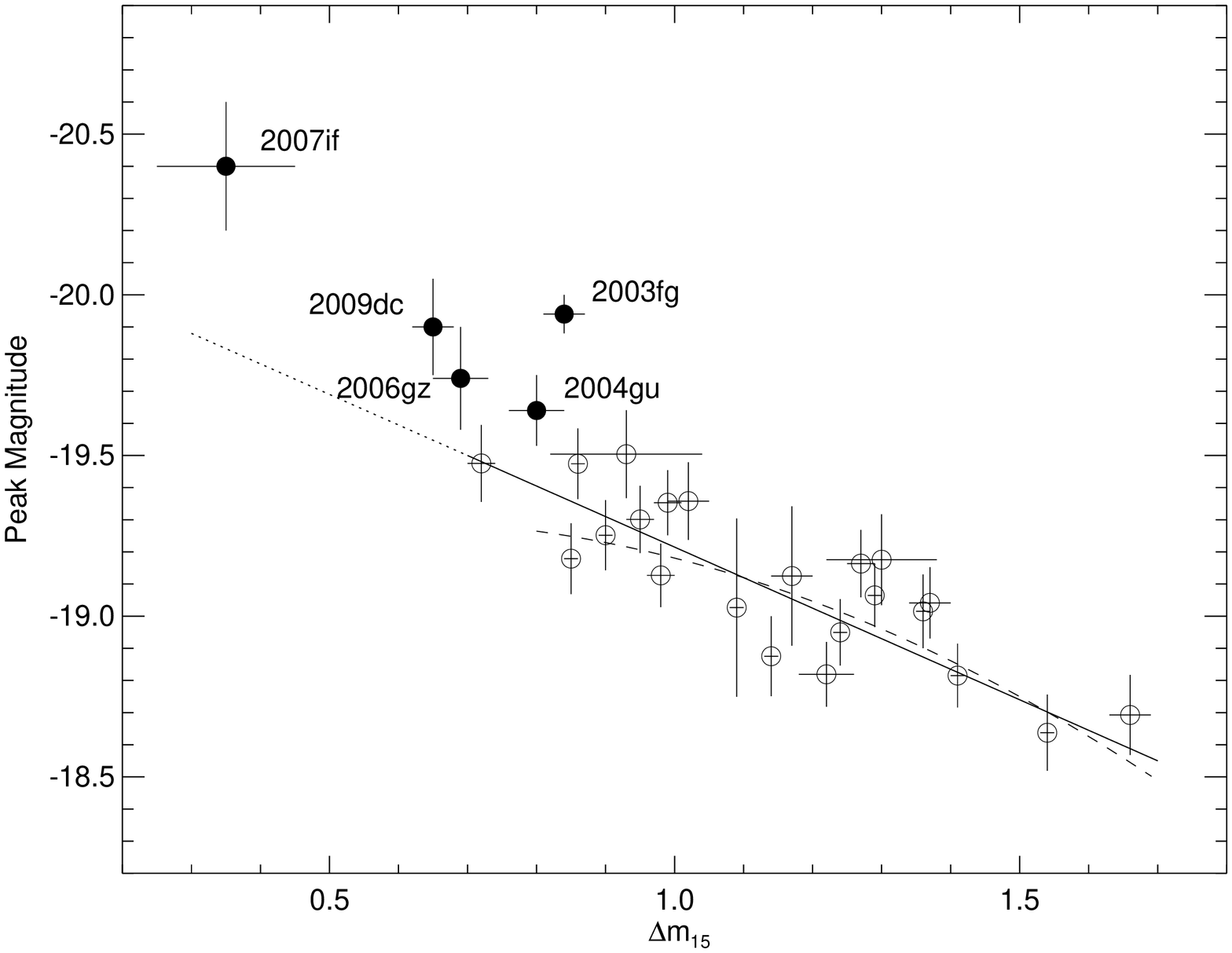,width=3.3in,angle=0}}
\caption{Peak luminosity vs. decay rate for SN 2007if and other SNe Ia. Except for SN 2007if, all the peak magnitudes are in V-band and the decay rates are measured or modeled in B-band. Data for the best observed sample in \citet{chpfs09} are plotted as open circles. SN 2007if and other over-luminous SNe Ia \citep{hsnec06, hgpbd07, ykkti09, chpfs09} are plotted as filled circles labeled with the name of each event. For SN 2003fg, the stretch factor of 1.13 is converted to $\Delta m_{15}$(B) using Eq (6) in \citet{pgggg97}. The magnitudes for SN 2007if, 2003fg and 2009dc are not corrected for host extinction. Host reddening for SN 2004gu is corrected in the same way as the other \citet{chpfs09}  events. The linear fit derived in \citet{chpfs09} and its extrapolation are over-plotted as a solid line and a dotted line respectively. The quadratic relationship in \citet{plssh99}, shifted to match the peak magnitude at  $\Delta m_{15}$(B)=1.1, is plotted as a dashed line. \label{fig:lum_width}}
\end{figure}

Given a reliable detection at 17.18$\pm$0.08 mag (corrected for the Galactic extinction of 0.21 magnitude in the r-band) by ROTSE-IIIb on September 5, SN 2007if, at a redshift 0.0736 (see $\S$2.3), reached about -20.4 absolute magnitude, without host reddening correction. This is one magnitude brighter than the average peak brightness of SNe Ia, even considering the zero-point uncertainty in our r-band calibration. As the supernova color evolves, the characteristic wavelength of the ROTSE response shifts. Direct comparison with narrow band light curves of other SNe Ia is thus problematic. We can nevertheless obtain approximate estimates of the characteristic parameters of the SN 2007if light curve. As measured by the ROTSE detections, the decay within 15 days from maximum light, $\Delta m_{15}$, is about 0.35 mag. This value is similar to the $\Delta m_{15}$(B) (decay in B-band) estimated from the relatively large post-maximum stretch factor \citep[][Eq~(6), although the obtained $\Delta m_{15}$ is outside the valid range of the relation]{pgggg97}. Given the typical small difference between the V and R-band maximum, we compare the peak luminosity and decay rate of SN 2007if with other SNe Ia in Figure~\ref{fig:lum_width}. Based on the relationship derived from fainter events \citep{pgggg97,plssh99,chpfs09}, SN 2007if was over-luminous by about a half magnitude.

\subsection{Spectroscopic Observations}

\subsubsection{Photometric Phase}
\begin{figure}[h]
\centerline{\psfig{file=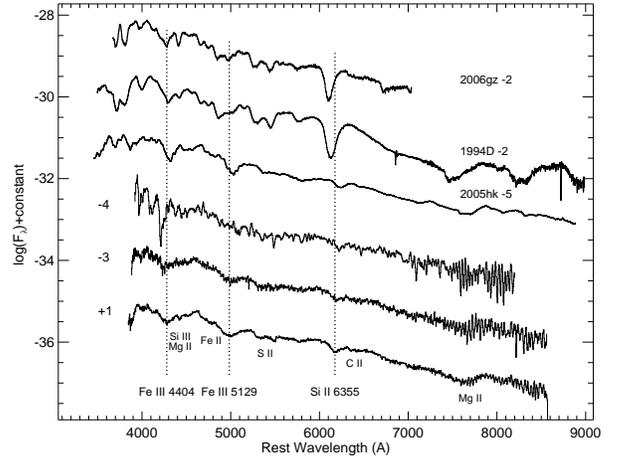,width=3.3in,angle=0}}
\caption{Early spectra of SN 2007if observed by HET, plotted against spectra at similar epochs for an over-luminous Ia, SN 2006gz \citep{hgpbd07}, a normal Ia, SN 1994D \citep{1996MNRAS.278..111P} and a peculiar Ia, SN 2005hk \citep{2007PASP..119..360P}. The spectra have not been corrected for host galaxy extinction. The vertical dotted lines mark Fe~III $\lambda$4404, $\lambda$5129 and Si~II $\lambda$6355 blueshifted by 8500~km/s\label{fig:2007if_spectra_early}}
\end{figure}

\begin{figure}[h]
\centerline{\psfig{file=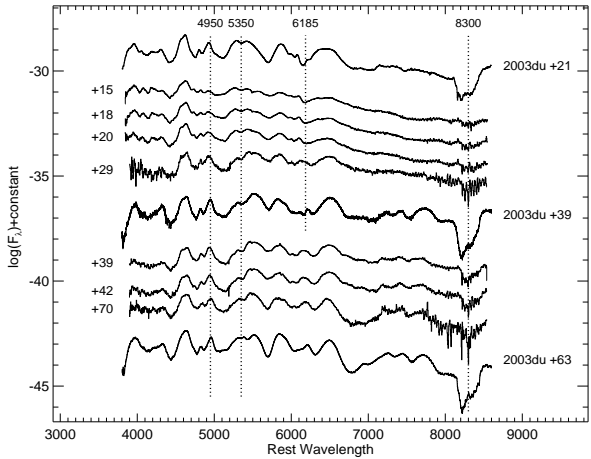,width=3.3in,angle=0}}
\caption{Spectra series of SN 2007if observed by HET, plotted against spectra at selected epochs for a normal Ia, SN 2003du \citep{2007A&A...469..645S}. \label{fig:2007if_spectra_late}}
\end{figure}

Ten spectra were obtained with the HET from Aug. 28 to Nov. 10, 2007 (see Table~\ref{tab:speclog}), covering 4 days before until 70 days after the estimated maximum light of Sep. 1.8 UT. These spectra were reduced using standard {\it IRAF} procedures. They are presented in Figure~\ref{fig:2007if_spectra_early} and Figure~\ref{fig:2007if_spectra_late} in comparison with other SN Ia spectra at similar epochs, obtained from the SUSPECT database \footnote{http://bruford.nhn.ou.edu/$\sim$suspect/}. The blue and red ends, where the spectra are dominated by noise, are truncated.


\begin{deluxetable}{cccccc}
  \tablecaption{Spectroscopic observations of SN 2007if \label{tab:speclog}}
  \tablewidth{0pt}
  \tablehead{\colhead{Age\tablenotemark{*}} & \colhead{MJD} & \colhead{Telescope} & \colhead{Wavelength} & \colhead{Expo.} &  \colhead{Reso-} \\
    \colhead{(days)} & \colhead{} & \colhead{/Instrument} & \colhead{Range ($\mbox{\AA}$)} & \colhead{Time (s)} &  \colhead{lution}}
  
  \startdata
  -4 & 54340.30 & HET/LRS & 4100-9200 & 1800 & 300 \\
  -3 & 54341.32 & HET/LRS & 4100-9200 &  900 & 300 \\
  +1 & 54345.31 & HET/LRS & 4100-9200 &  900 & 300 \\
 +15 & 54359.44 & HET/LRS & 4100-9200 &  900 & 300 \\
 +18 & 54362.44 & HET/LRS & 4100-9200 &  900 & 300 \\
 +20 & 54365.23 & HET/LRS & 4100-9200 &  900 & 300 \\
 +29 & 54374.22 & HET/LRS & 4100-9200 &  900 & 300 \\
 +39 & 54384.18 & HET/LRS & 4100-9200 &  900 & 300 \\
 +42 & 54387.18 & HET/LRS & 4100-9200 &  900 & 300 \\
 +70 & 54414.30 & HET/LRS & 4100-9200 &  900 & 300 \\
 +339& 54683.58 & Keck/LRIS & 3400-9200 & 1800 & 1000 \\
 +688\tablenotemark{$\dagger$}& 55032.60 & Keck/LRIS & 3100-10000 & 4200 & 1000 \\
  \enddata
  \tablenotetext{*}{Relative to the estimate maximum light of Sep. 1.8 UT.}
  \tablenotetext{$\dagger$}{Observation of the host.}
\end{deluxetable}

The spectrum taken at about four days before maximum had relatively low signal to noise and is smoothed for display in Figure~\ref{fig:2007if_spectra_early}. It shows a blue continuum with a probable broad and shallow absorption between 4700 and 6000~$\mbox{\AA}$ but no distinguishable features that can be associated with a SN Ia. At day -3, shallow P-Cygni profiles are visible around 4300, 5000 and 6100~$\mbox{\AA}$, consistent with the Fe~III multiplets around $\lambda$4404 and $\lambda$5129 and Si~II $\lambda$6355. These defining features of a SN Ia become more evident in the spectrum taken just past maximum, at day +1. The lack of O~I $\lambda$7773 argues against SN 2007if being a Type Ic. Contributions from Si~III, Mg~II, Fe~II and S~II can also be identified in the day +1 spectrum along with a probable trace of C~II (see Figure~\ref{fig:2007if_spectra_early}). Although the C~II feature is uncertain, it has been detected in other luminous SNe Ia before or around maximum light \citep{hsnec06,hgpbd07,ykkti09}.

Shallow Si~II features are usually seen in luminous SNe Ia (i.e. the 1991T-like class) before and around maximum. For SN 2007if, the Fe signatures are not particularly strong. The relative strength of Fe seems to be between a normal SN Ia and a 1991T-like event. The shallow features may be caused by high temperatures in the outer layers that ionize Si~II and Fe~II \citep{mdt95,npbbh95}. Indeed, Figure~\ref{fig:2007if_spectra_early} shows that the spectral shape of SN 2007if is similar to that of the pre-maximum SN 2005hk \citep{2007PASP..119..360P}, which displayed significantly less blueshifted absorptions and was a peculiar sub-luminous event with a high initial temperature. Another probable explanation for shallow features is dilution by an underlying continuum emission. This prospect will be further discussed in $\S$3.

The spectroscopic resemblance of SN 2007if to SN 2003fg was pointed out in \citet{atel1212}, where a spectrum of 2007if taken on Sep. 10.5 by the SNfactory was reported to closely match that of SN 2003fg at 2 days post-maximum. We note that at day +1, closer to the phase when SN 2003fg was observed, the features have not yet grown to be as strong as seen in SN 2003fg, probably suggesting a even higher temperature or more dilution in SN 2007if.

The expansion velocity measured from the Si~II $\lambda$6355 absorption minimum at 1 day post-maximum is 8500$\pm$400 km/s. This photospheric velocity is confirmed if Fe~II $\lambda$4404 and Fe~II $\lambda$5129 are mainly responsible for the minima of the other two major absorption features on the blue side. Such a velocity, clearly slower than in a typical SN Ia (Figure~\ref{fig:2007if_spectra_early}) and much slower than for a luminous 1991T-like event, is consistent with that measured for SN 2003fg at a similar epoch around maximum \citep{hsnec06}. This further strengthens the proposition that SN 2007if is in the same category as 2003fg. Another over-luminous event, SN 2006gz, however, does not have shallow Si~II features and has a typical photospheric velocity.

Between day +15 and day +70, the spectral evolution of SN 2007if closely resembles that of a normal SN Ia during similar epochs (e.g. SN 2003du \citealp{2007A&A...469..645S}, a normal Type Ia with good spectroscopic coverage, see Figure~\ref{fig:2007if_spectra_late}). Overall, the spectra of SN 2007if have less contrast than those of 2003du. One remarkable deviation is the weak features from intermediate mass elements in SN 2007if. The relatively slow Si~II $\lambda$6355 absorption disappears into a blend with Fe~II lines by day +29, while its core is often visible until day +40 or later for normal SN Ia. The Ca~II IR triplet can be identified at around 8300$\mbox{\AA}$ after day +15, but is also considerably weaker than typically observed.

Another noticeable difference is at around 4950$\mbox{\AA}$, where the spectra of SN 2007if appear comparatively flat before day +20. The peak at 4950$\mbox{\AA}$ grows stronger over time and becomes indistinguishable from a typical SN Ia by day +29. Such a trend, together with the rapidly decreasing blue-to-red peak ratio around 5350$\mbox{\AA}$, is observed for SN 1994D from 2 to 4 weeks post-maximum \citep[see Figure~7 in ][]{2005PASP..117..545B} and was modeled as due to strengthening Fe II and Cr II lines.

The photospheric velocity evolution is not constrained from the Si~II $\lambda$6355 absorption as it is not resolved in our later observations, but we note that the other absorption minima show similar line velocities to normal events. The inner layers of the SN 2007if ejecta thus do not necessarily have low kinetic energy.

\subsubsection{Nebular Phase}
\begin{figure}[h]
\centerline{\psfig{file=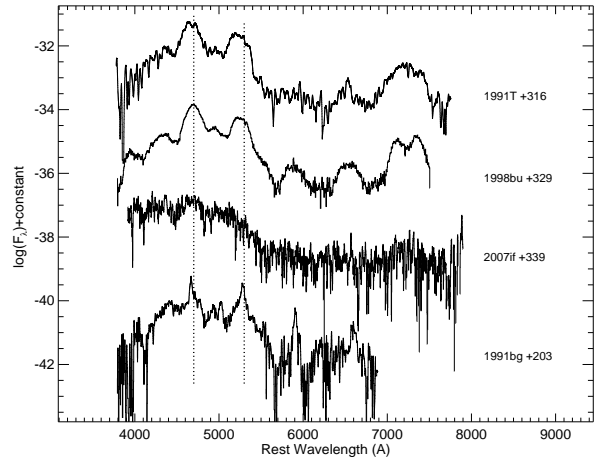,width=3.3in,angle=0}}
\caption{Nebular spectrum of SN 2007if observed by Keck, smoothed and plotted against spectra at similar epochs for the luminous SN 1991T \citep{1998AJ....115.1096G}, normal SN 1998bu \citep{2001ApJ...549L.215C} and the sub-luminous SN 1991bg \citep{tbcdd96}. \label{fig:2007if_spectra_nebular}}
\end{figure}

SN 2007if was observed in the nebular phase by Keck on Aug. 5, 2008. At day +339, the spectrum shows a broad bump on the blue side and a probable broad emission feature just above 7000$\mbox{\AA}$. While the general shape of the spectrum is similar to that of other SNe Ia in nebular phase (see Figure~\ref{fig:2007if_spectra_nebular}), the typical narrow emission features at 4700 and 5300$\mbox{\AA}$ are not as prominent. These two features are usually modeled as dominated by [Fe III] and a combination of [Fe II] and [Fe III] forbidden lines in the blue and red respectively \citep{1973ApJ...185..303K, axe80,1992ApJ...400..127R}. The slightly higher ratio of flux density at these two wavelengths may indicate a somewhat higher temperature in the inner layers of SN 2007if than a normal SN Ia.

Further quantitative analysis is complicated by the contamination from the host galaxy. We find that the total SN plus host galaxy flux as observed in the g-band by Keck/LRIS on Aug. 4, 2008 was only 0.4 mag brighter than the host flux alone (see $\S$2.3). As estimated from the difference, SN 2007if was at about 23.9$\pm$0.4 mag in g-band. As for the featureless pre-maximum spectra, dilution by an additional continuum may also play a role. 

A similarly featureless nebular spectrum in the blue region was observed for over-luminous SN 2006gz \citep{2009ApJ...690.1745M}. However, in the latter case, relatively strong emission was detected at $\sim$7200-7300$\mbox{\AA}$ (probably due to [Ca II]), indicating yet different composition and conditions. 

\subsection{Observations of the Host Galaxy}
The host of SN 2007if was imaged by Keck on June 26.6, 2009, long after the SN had faded away. An R-band magnitude of 22.70$\pm$0.08 and g magnitude of 23.03$\pm$0.13 (corrected for Galactic extinction) were estimated by calibrating to the SDSS measurements of nearby objects. Based on the g-R color, the host of SN 2007if is among the bluest objects within the 3\arcmin\ neighborhood.

\begin{figure}[h]
\centerline{\psfig{file=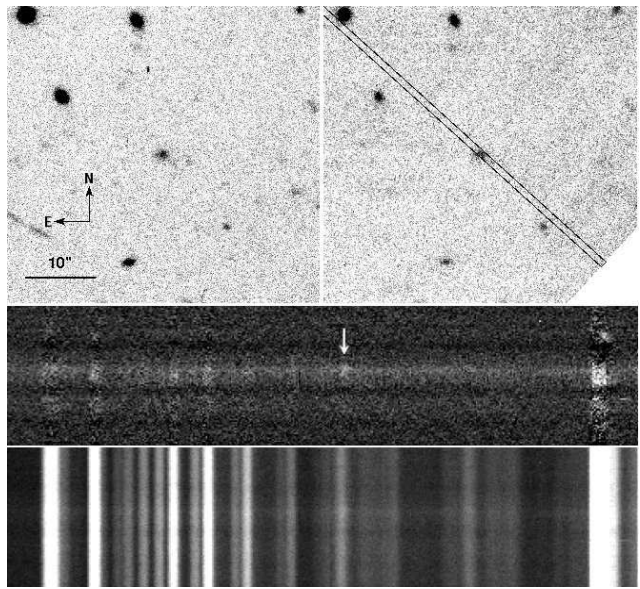,width=3.3in,angle=0}}
\caption{Optical images and spectra of the host taken by Keck/LRIS two years after the SN discovery. The top two panels show the R-band (left) and unfiltered blue channel (right) images, with the slit position marked in the latter. The middle panel shows the sky subtracted two-dimensional spectrum. The remaining emission feature, identified as H$_{\alpha}$, is marked with a downward arrow. The original spectrum, dominated by strong sky features, is displayed in the bottom panel. \label{fig:2007if_host_2d}}
\end{figure}

A spectrum of the host was also obtained by Keck on July 20.6, 2009. Because of the faintness of the host, the spectrum was dominated by sky emissions. After removing the sky lines, a single emission feature was identified as H$_{\alpha}$ at redshift 0.0736 (Figure~\ref{fig:2007if_host_2d}). This redshift is fully consistent with the estimation from cross-correlating the spectra with templates in SNID \citep{2007ApJ...666.1024B}. We therefore adopt this as the redshift of SN 2007if.

\begin{figure}[h]
\centerline{\psfig{file=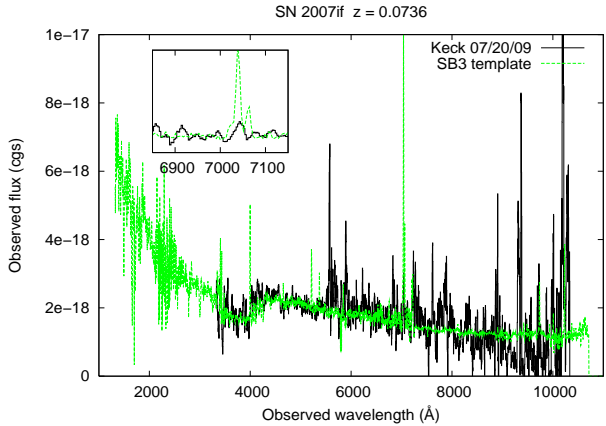,width=3.3in,angle=0}}
\caption{Host spectrum of SN 2007if, truncated on the noisy blue side and plotted against a SB3 type galaxy template at redshift 0.07. \label{fig:2007if_host_keck07}}
\end{figure}

At the measured redshift, the host has an absolute R magnitude of -14.91. The host spectrum is best-matched by an SB3 - SB4 galaxy template at redshift 0.07 using ``superfit'' \citep{2005ApJ...634.1190H}, but the H$_{\alpha}$ emission, indicative of massive stars and recent star formation activity, is much weaker than that in the template (see Figure~\ref{fig:2007if_host_keck07} and the inset).

\section{Discussion}
If the supernova light curve is entirely powered by radioactive decay of $^{56}$Ni, the luminosity of the supernova at maximum light would roughly equal the instantaneous energy deposition from $^{56}$Ni. The peak brightness of the supernova thus provides an estimate of the amount of $^{56}$Ni synthesized in the explosion \citep{1982ApJ...253..785A}.

A precise estimate of the bolometric luminosity is not possible without multi-band photometry. Since the observed evolution is quite similar for SN 2007if and a normal brightness SN Ia, we can simply scale its $^{56}$Ni production with the peak luminosity. At one magnitude brighter than average, SN 2007if would require about 1.5~\msun\ of $^{56}$Ni compared to a typical 0.6~\msun\ \citep{2000A&ARv..10..179L}. If the weak features from intermediate mass elements indicate a more complete burning, the total mass requirement is less extreme. A larger fraction of $^{56}$Ni production is consistent with a trend suggested by modeling of the nearby ``normal'' SNe Ia \citep{2007Sci...315..825M}. Even considering the large uncertainty associated with such an estimate, the amount of $^{56}$Ni is far beyond that which can be synthesized in a 1.4~\msun\ white dwarf progenitor.

\citet{hsnec06} discussed a large progenitor mass and hence a large binding energy as the explanation for the apparent discrepancy between the exceptional brightness and a low photospheric velocity of SN 2003fg. A similar situation is noticed for SN 2007if. In addition, we observe shallow features in the early spectra, suggesting high ionization and high temperature. At later times, the spectral evolution of SN 2007if follows that of a typical SN Ia, indicating similar physical processes.
 
Despite the resemblance to SN 2003fg, the spectra of SN 2007if clearly differ from that of another ultra-luminous SN Ia with a proposed super-Chandrasekhar progenitor, 2006gz (Figure~\ref{fig:2007if_spectra_early}). For the latter case, prominent narrow absorption features were observed from 2 weeks before maximum light. \citet{2009MNRAS.394..239M} investigated the different properties between SN 2003fg and 2006gz, and suggested that they might be from the same type of progenitor but observed at different angles. They propose that the shallow spectral features can be explained as due to less shocked material at the pole. However, polarimetric observations of a recent luminous SN Ia, 2009dc, which has similar observational features as SN 2003fg, do not support the aspherical explosion hypothesis \citep{2009arXiv0908.2057T}.

As an alternative, we consider the interaction between the ejecta and the circumstellar medium (CSM) that may produce extra luminosity. Detection of a CSM signature may provide important clues about the mass transfer process leading to the SN Ia explosion. A CSM played a domninant role in SN 2002ic \citep{2003Natur.424..651H,2004ApJ...604L..53W} that had an underlying SN Ia spectrum, but a strong (asymmetric) Type IIn CSM interaction. SN 2006X \citep{2007Sci...317..924P}, SN 1999cl \citep{2009ApJ...693..207B} and SN 2007le \citep{2009ApJ...702.1157S} showed variable Na D lines that indicated the presence of a more dilute CSM. Many SN Ia show high-velocity Ca II features \citep{wbhkw03,2004ApJ...607..391G,2005ApJ...623L..37M} that may hint at a CSM. Perhaps ``super Chandrasekhar'' events reveal excess luminosity from CSM interaction. In such a scenario, the continuum emission above the photosphere produces a ``toplighting'' effect \citep{2000PASP..112..217B}. As modeled in \citet{2000PASP..112..217B}, when the external light has comparable intensity as the photospheric emission, the spectral features are strongly ``muted.'' The shallow features observed in SN 2007if could then be a consequence of luminous emission from circumstellar interaction, roughly doubling the total luminosity. It is potentially possible that deceleration by the reverse shock results in shifted absorption minima and apparently low photospheric velocity. Any CSM interaction must satisfy the constraint that narrow emission lines are not observed in SN 2007if. In addition, any CSM cannot correspond to a steady-state wind since an r$^{-2}$ density profile would generate substantial luminosity at early times and alter the light curve in a manner that is not observed \citep[see e.~g. Fig.~5 in ][]{2004ApJ...607..391G}. Such a CSM might be reminiscent of the expanding gas identified in novae by \citet{2008ApJ...685..451W} or a common envelope in a DD model \citep{kmh93}. It is not clear how the interaction can be fine tuned to make a contribution comparable to the nuclear decay throughout the SN evolution. Modeling of the exact distribution of the CSM and the propagation of forward/reverse shocks is beyond the scope of this paper.

After our paper was submitted, we became aware of the independent analysis of \citet{scalzo10}. \citet{scalzo10} estimate a similar $^{56}$Ni mass from the bolometric light curve and argue the low photospheric velocity is a likely result of the interaction between the ejecta and an massive envelope surrounding a double-degenerate progenitor.

\section{Conclusions}
SN 2007if was one of the first discoveries of the ROTSE Supernova Verification Project (RSVP) \citep{ya08}, which extends the efforts of the Texas Supernova Search (TSS, \citealp{quimby_thesis}) and uses all four ROTSE-III telescopes. 

The peak unfiltered magnitude of SN 2007if (calibrated to SDSS r-band) measured by ROTSE-IIIb, corresponds to an absolute magnitude of -20.4 at redshift 0.074. This is by far the brightest SN Ia observed. If powered by $^{56}$Ni decay, the exceptional brightness of SN 2007if requires a total $^{56}$Ni mass close to or even exceeding the Chandrasekhar mass of a non-rotating white dwarf. The photospheric expansion velocity derived around maximum is comparable to the estimate for SN 2003fg and suggests a relatively low kinetic energy. After two weeks post-maximum, the spectral evolution of SN 2007if become indistinguishable from a normal SN Ia, except for ``muted'' features and weak absorptions due to Si and Ca. Late time observations show that SN 2007if occured in a low luminosity host, similar to that of SN 2003fg. 

The observed properties of SN 2007if may suggest that it had a massive progenitor, well above 1.4~\msun, as proposed for a few other luminous SNe Ia. Super-Chandrasekhar explosions have been investigated for systems with rapid rotations \citep{unh03,yl05}. Both a single WD or a merger of two degenerate stars can sustain a larger mass if rapidly rotating.

CSM interaction may provide another interpretation for the excess luminosity. In the presence of such emission above the photosphere, shallow spectral features are predicted, consistent with our observation of SN 2007if. However, it is noted that the spectra of other luminous SNe, such as 2003fg and 2006gz are not similarly ``muted.'' Further studies are needed to see whether the observed light curve can be reproduced.

So far, the few most luminous SNe Ia show diverse behaviors. In general, the events follow a positive correlation between peak brightness and stretch factor, except for SN 2003fg \citep{hsnec06}, which was among the brightest but showed a relatively fast decay. SN 2009dc \citep{ykkti09} and 2007if had similarly slow photospheric velocity derived from Si~II 6355 as for SN 2003fg. SN 2006gz \citep{hgpbd07} and 2004gu \citep{chpfs09} appear as close cousins, with relatively high photospheric velocity. Suppression of the silicon velocity was also noticed at early times for SN 2006gz \citep{hgpbd07}. Although rare, over-luminous SNe Ia show significant deviations in brightness when the normal parametrization method is applied in cosmological studies \citep{fpbch10}. More luminous events, polarimetric observations, X-ray monitoring and detailed simulations in the future may help to understand the category as a whole. 

\acknowledgments
The authors gratefully acknowledge M. Kasliwal for the data taken at Keck, and C.~S.~Peters and J.~R.~Thorstensen for the data taken at the 2.4m Hiltner telescope. Special thanks to the McDonald Observatory staff and the H.E.S.S. staff, especially David Doss and Toni Hanke. We would also like to thank the staff of Hobby-Eberly Telescope for their support. FY is supported by the NASA {\it Swift} Guest Investigator grants NNX-08AN25G. JCW is supported in part by NSF AST-0707769. ROTSE-III has been supported by NASA grant NNX-08AV63G, NSF grant PHY-0801007, the Australian Research Council, the University of New South Wales, the University of Texas, and the University of Michigan. The authors thank the anonymous referee for the valuable comments.




\end{document}